# Thermal-noise-limited, compact optical reference cavity operated without a vacuum enclosure

Yifan Liu[1,2], Charles A. McLemore[1,2], Megan Kelleher[1,2], Dahyeon Lee[1,2], Takuma Nakamura[1,2], Naijun Jin[3], Susan Schima[2], Peter Rakich[3], Scott A. Diddams[1,2,4] and Franklyn Quinlan[1,2,4]

1. Department of Physics, University of Colorado Boulder, 440 UCB Boulder, CO 80309, USA
2. Time and Frequency Division, National Institute of Standards and Technology, Boulder, CO 80305, USA
3. Department of Applied Physics, Yale University, New Haven, CT 06520, USA
4. Electrical, Computer and Energy Engineering, University of Colorado Boulder, 425 UCB Boulder, CO 80309, USA
*yifan.liu@colorado.edu, franklyn.quinlan@nist.gov*

*Abstract*—We present an in-vacuum bonded, 9.7 mL-volume Fabry-Pérot ultrastable optical reference cavity that operates without a vacuum enclosure. A laser stabilized to the cavity demonstrates low, thermal noise-limited phase noise and $5 \times 10^{-14}$ Allan deviation at 1 second.

## I. Introduction

Cavity-stabilized CW lasers are important components in various applications, such as optical clocks and optically derived low-noise microwaves. Although state-of-the-art cavity-stabilized laser systems can achieve better than $10^{-16}$ fractional frequency instability [1, 2], the Fabry-Pérot (FP) cavity used in such systems are either longer in length or operating in cryogenic temperatures, making the overall system bulky and not conducive to out-of-laboratory applications. Applications such as portable optical clocks [3], portable microwave sources [4], and environmental sensing [5], demand more compact, lightweight, and simplified cavity systems. Recently, compact optical reference cavities have demonstrated sub-$10^{-13}$ level fractional frequency instability with a cavity volume less than 10 mL [6, 7]. However, to maintain such excellent performance, these cavities still need to operate in a high-vacuum (HV) environment. The HV environment typically consists of a vacuum chamber, an ion pump and an HV valve for the initial pumping process, resulting in a combined size much larger than that of the small cavity. To make full use of the achievable small cavity size and to simplify use, here we demonstrated an in-vacuum-bonded compact optical reference cavity that does not require any vacuum enclosure to achieve near thermal-noise-limited performance and $10^{-14}$ level fractional frequency instability. Importantly, no degradation of performance has been observed in the 4.5 months since the cavity was bonded.

The optical reference cavity we built has a cylindrical shape with a diameter of 25.4 mm and an overall length of 19.05 mm. It is made of ultra-low expansion (ULE) glass, consisting of two 6.35 mm long cylindrical mirror substrates, one flat and one curved (1 m radius of curvature), each with a highly reflective dielectric coating, and a 6.35 mm long cylindrical spacer with a center bore hole in between (Fig. 1b). The superpolished surfaces on the mirrors and spacer have an average surface roughness of about 1Å, smooth enough for optical contact bonding. The three components are optical contact bonded inside a vacuum bonder we designed, as illustrated in Fig. 1a. The vacuum bonder consists of an HV bellows valve and a specially designed vacuum-compatible holding structure that aligns the three components in place. To bond the cavity, we put one mirror substrate at the bottom of the holding structure and put the spacer directly on that substrate. The top part of the holding structure acts as a clamp that holds the remaining mirror substrate ~2 mm above the top surface of the spacer by friction. After putting the holding structure containing all three components into the HV valve, we seal the valve and pump out the air with a turbo pump. The pressure is monitored by a

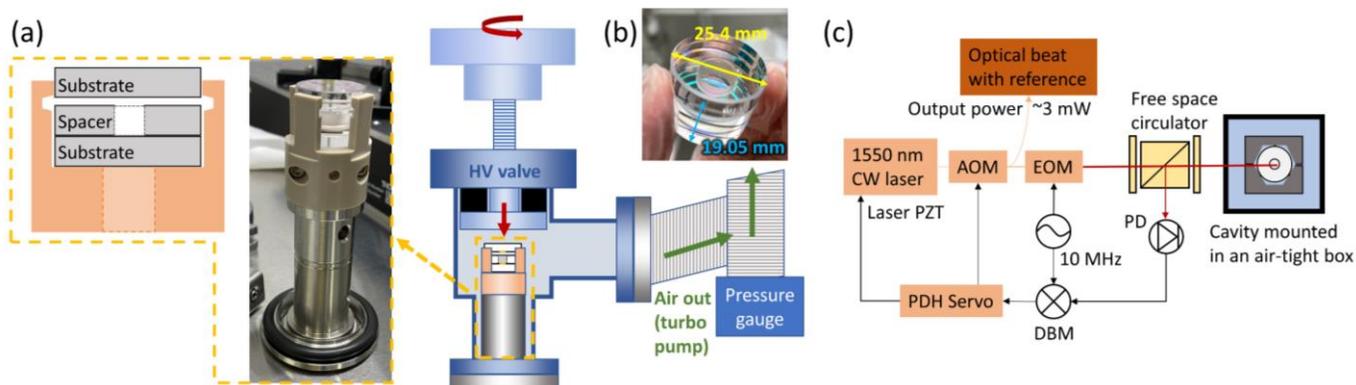

Fig. 1. (a) Schematic illustration and photo of the vacuum bonder design. (b) Photo of the in-vacuum bonded cavity. (c) Simplified diagram of PDH locking setup. AOM, acousto-optic modulator; EOM, electro-optic modulator; DBM, double-balanced mixer; PD, photodetector.

This work is funded by DARPA and NIST. Product of the US government, not subject to copyright in the USA.

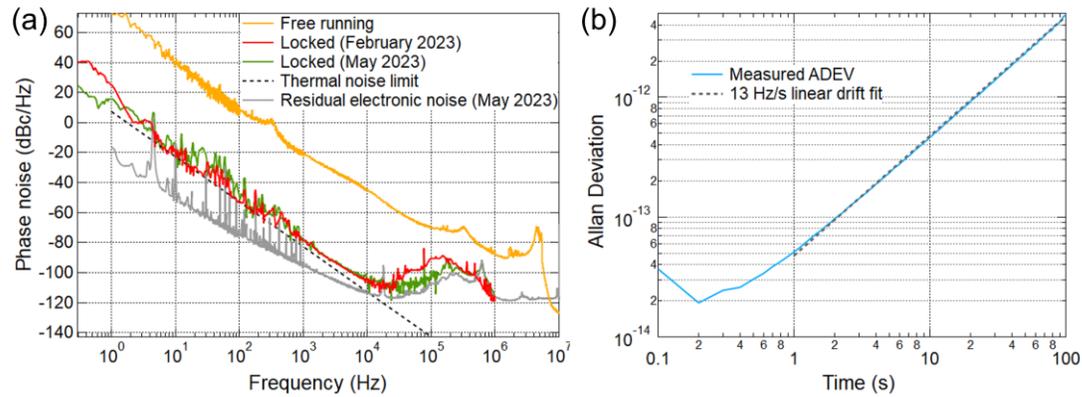

Fig. 2. (a) Phase noise measurement of the 1550 nm laser PDH locked to the in-vacuum bonded cavity performed two weeks (in February 2023) and 4 months (in May 2023) after initial bonding. (b) Allan deviation of the 1550 nm laser PDH locked to the in-vacuum bonded cavity. No degradation in performance has been observed during the ~4 months after initial bonding.

pressure gauge. Once the pressure inside the valve is lowered to $10^{-3}$ Pa, we rotate the handwheel to lower the inner top surface of the valve such that it presses the top mirror substrate to be in contact with the spacer surface and applies constant pressure for the surfaces to bond. We then shut down and remove the turbo pump. The bonded cavity is held under a continuous compression force, surrounded by atmosphere, for two days. Due to the high surface quality of the spacer and mirrors, we expect the optical contacting bond to be largely air-tight, sealing vacuum inside the cavity. In this way, we are able to create a cavity that is itself a mini-vacuum chamber and can thus maintain good stability of its intra-cavity optical path length, resulting in good frequency stability without a bulky vacuum enclosure.

## II. RESULTS

The in-vacuum bonded cavity is measured to have a finesse of 850,000 at 1550 nm, corresponding to a linewidth of 27.8 kHz. To test the performance of the cavity, we mount the cavity into a V-groove holder enclosed by an air-tight stainless-steel box for better temperature and outer air pressure control, as well as acoustic isolation. The air-tight box is not evacuated, and its temperature is actively maintained near room temperature. A 1550 nm commercial fiber laser is stabilized to the cavity via the Pound-Drever-Hall (PDH) method (Fig. 1c). The laser is PDH locked to a 1550 nm TEM00 mode of the cavity with fast feedback control achieved through an acousto-optic modulator (AOM) and slow feedback control through the laser's piezoelectric tuning port. In order to probe the fundamental limits of our cavity, the system is placed within a passive acoustic isolation box with active vibration cancellation.

To measure the phase noise of the cavity-stabilized light, we split the stabilized light into two channels to compare with two different optical references, an optical frequency comb (OFC) stabilized to a local oscillator of a Yb clock and a 1550 nm cavity stabilized laser that is offset by 8.9 GHz from our cavity light. The two heterodyne beat notes with the two references are digitally sampled and cross-correlated to reveal the noise of our cavity stabilized laser. As shown in Fig. 2a, the measured phase noise of the laser follows the estimated thermal noise limit from 10 Hz to 1 kHz offset and reaches around -105 dBc/Hz at 10 kHz offset. The phase noise can be used to estimate the residual gas level in the cavity [8], for which we determine an upper limit of 500 Pa. Thus, near-thermal noise limited phase noise performance can be achieved under only modest vacuum levels. To measure the fractional frequency stability, we track the heterodyne beat between our light and the stabilized OFC to acquire the laser's fractional frequency change over time and calculate Allan deviation (ADEV). As shown in Fig. 2b, the ADEV at 1 s reaches $5 \times 10^{-14}$ and is dominated by a ~13 Hz/s linear drift at longer timescales.

In summary, we demonstrated a method for in-vacuum bonding of a FP optical reference cavity such that it can operate without the need for a high-vacuum enclosure. Near thermal noise-limited optical phase noise performance and fractional frequency instability of $5 \times 10^{-14}$ at 1 s are achieved with only an air-tight box surrounding the cavity. Our cavity bonding method is compatible with other cavity geometries, such as long spacer length and smaller cross-sectional diameter. These results point to a new path toward compact and portable ultrastable laser systems for out-of-lab applications.